\documentclass[prd,twocolumn,showpacs,preprintnumbers,nofootinbib,amsmath,amssymb,floatfix]{revtex4}

\usepackage{graphicx}
\usepackage{dcolumn}
\usepackage{bm}
\usepackage{amsmath}

\newcommand{\beq}{\begin{eqnarray}}
\newcommand{\eeq}{\end{eqnarray}}

\newcommand{\real}{{\sf I}\kern-.12em{\sf R}}
\newcommand{\comp}{{\sf I}\kern-.50em{\sf C}}
\newcommand{\unity}{{\sf I}\kern-.54em{\sf 1}}

\def\spose#1{\hbox to 0pt{#1\hss}}
\def\ltapprox{\mathrel{\spose{\lower 3pt\hbox{$\mathchar"218$}}
 \raise 2.0pt\hbox{$\mathchar"13C$}}}

\begin{document}

\title{The critical line of two-flavor QCD at finite isospin or baryon 
densities from imaginary chemical potentials}
\author{Paolo Cea}
\affiliation{Dipartimento di Fisica dell'Universit\`a di Bari and INFN - Sezione di Bari, 
I-70126 Bari, Italy}
\email{paolo.cea@ba.infn.it}
\author{Leonardo Cosmai}
\affiliation{INFN - Sezione di Bari, I-70126 Bari, Italy}
\email{leonardo.cosmai@ba.infn.it}
\author{Massimo D'Elia}
\affiliation{
Dipartimento di Fisica dell'Universit\`a
di Pisa and INFN - Sezione di Pisa,\\ Largo Pontecorvo 3, 56127 Pisa, Italy}
\email{delia@df.unipi.it}
\author{Alessandro Papa}
\affiliation{Dipartimento di Fisica dell'Universit\`a della Calabria \\
and INFN - Gruppo collegato di Cosenza, 
I-87036 Arcavacata di Rende, Cosenza, Italy}
\email{papa@cs.infn.it}
\author{Francesco Sanfilippo}
\affiliation{Laboratoire de Physique Th\'eorique (Bat. 210) Universit\'e Paris
SUD, F-91405 Orsay-Cedex, France \\
and INFN, Sezione di Roma, Piazzale Aldo Moro 5, I-00185 Roma, Italy}
\email{francesco.sanfilippo@th.u-psud.fr}

\date{\today}

\begin{abstract}
We determine the (pseudo)critical 
lines of QCD with two degenerate staggered fermions at nonzero temperature and 
quark or isospin density, in the region of imaginary 
chemical potentials; analytic continuation is then used
to prolongate to the region of real 
chemical potentials.
We obtain an accurate determination of 
the curvatures at zero chemical potential,
quantifying the deviation between the case of finite quark and of 
finite isospin chemical potential.
Deviations from a quadratic dependence 
of the pseudocritical lines
on the chemical potential are clearly seen in both cases:
we try different extrapolations and, 
for the case of nonzero isospin 
chemical potential, confront them 
with the results of direct Monte Carlo simulations. 
Finally we find that, as for the finite quark density case,
an imaginary isospin chemical potential can strengthen
the transition till turning it into strong first order.
\end{abstract}


\maketitle

\section{Introduction}
\label{introd}

The determination of the QCD phase diagram in the temperature
- quark density plane is becoming increasingly important, due to its
impact in cosmology and in the physics of compact stars and of
heavy-ion collisions.

The first-principle nonperturbative approach of discretizing QCD on a 
space-time lattice and performing numerical Monte Carlo simulations 
is plagued, at nonzero quark chemical potential, 
by the well-known sign problem:
the fermionic determinant is complex and the Monte Carlo sampling becomes
unfeasible.
An exact solution to the problem is yet not known, but various 
approximate
alternatives have been explored. One possibility is to investigate
other models which are free of the sign problem, like two-color QCD 
or QCD with a finite isospin density~\cite{son,kogut}; however in such approach
predictivity is plagued by the fact that systematic differences between 
the model and the original theory are not known {\it a priori}. 

A positive measure is obtained also when the quark chemical potential
is purely imaginary, in this case the idea is to infer
the behavior 
at real chemical potential by analytic continuation.
Such approach was first suggested 
in Ref.~\cite{Alford:1998sd}, while the effectiveness of the method
of analytic continuation was pushed forward in Ref.~\cite{Lombardo:1999cz}.
Since then, the method has been extensively applied to  
QCD with staggered~\cite{muim,immu_dl,azcoiti,chen,defor06,sqgp,2im,cea3} 
and Wilson quarks~\cite{Wu:2006su,NN2011} and tested in QCD-like theories 
free of the sign problem~\cite{Hart:2000ef,giudice,cea,cea1,conradi,Shinno:2009jw,cea2} and in spin models~\cite{potts3d,kt}. 

The idea underlying the method of analytic continuation 
is very simple: if the dependence of an 
observable or of the critical line itself on the imaginary quark chemical 
potential is expressed in terms of an analytic function inside a certain 
domain, then this analytic function can be prolongated to the largest possible
domain, compatible with the presence of singularities, up to the physically 
relevant region of real chemical potentials. 

There are, however, two important limitations to the effectiveness of the 
method: a practical one, due to the fact that Monte Carlo simulations 
yield data points (with statistical uncertainties) at fixed values of the 
imaginary chemical potential and, therefore, analytic continuation passes
through the choice of an interpolating function, which may be ambiguous;
a principle one, due to the nonanalyticities and the periodicity of the theory 
with imaginary chemical potential~\cite{rw},  which makes so that the region 
effectively available for Monte Carlo simulations is limited by the condition 
Im$(\mu)/T\lesssim 1$. The combination of these two drawbacks implies that the 
region of real chemical potentials where the analytic continuation is expected 
to be reliable can be estimated as Re$(\mu)/T\lesssim 1$.

Apart from analytic 
continuation, a careful study of the phase diagram
in the $T$ - Im($\mu$) plane is important by its own.
It may teach us something about the critical properties of
QCD also at zero or small real 
$\mu$~\cite{defor06,sqgp,deph2,rwep,rwep_nf3,rwep2,rwep3}
and, at the same time, it can be used to the test the reliability of QCD-like
models which are then used to explore real $\mu$ as
well~\cite{kouno,braun,sakai,aarts,pawlowski,kouno_2011,rafferty,pagura,kashiwa,morita}.

An important role is played
in this respect by the periodic series of unphysical first order lines,
located at Im$(\mu)/T = (2 k + 1)\pi/N_c$, { $k=0,1,2,\ldots$,} 
$N_c$ being the number of colors, 
which characterize the high-$T$ region of the $T$ - Im($\mu$) plane
(Roberge-Weiss lines). Such lines are connected to the analytic 
continuation of the physical pseudocritical line by an endpoint
which, both for $n_f = 2$ and $n_f = 3$ QCD, 
is  first order (triple point) 
in the limit of small or high quark masses and second
order for intermediate mass values; 
it has been conjectured that this may be directly 
related to the nature of the phase transition 
at zero or small real chemical potential~\cite{rwep,rwep_nf3,rwep2,rwep3}.

In a series of studies by some of us, we have started a detailed 
investigation aimed at checking the reliability 
of analytic continuation of the pseudocritical line and at 
extending its range of applicability, by looking for possible deviations
from the simple linear behavior { in $\mu^2$,} 
$T_c(\mu^2) = T_c(0) + A \mu^2$, which fitted well 
with earlier studies~\cite{muim,immu_dl}.
To this purpose, the range of Im$(\mu)$ values 
included in numerical simulations was extended with respect to earlier
studies, up to reaching the border
of the first Roberge-Weiss (RW) sector, 
statistics was considerably increased 
and different kind of interpolating functions were considered. 
In order to
validate the different interpolation options, in some cases QCD-like theories
were adopted (such as SU(2) or SU(3) with finite isospin density) which, 
being free of the sign problem, allowed the comparison 
of extrapolations with the results of Monte Carlo simulations performed
directly at real $\mu$. A detailed summary of such investigation is 
reported in Section~\ref{sectionII}, the main result emerging from it 
is that non-linear corrections are not negligible, but
an unambiguous extrapolation to real $\mu$ fails for 
$\mu/T \sim O(1)$.

In the present study we consider $n_f = 2$ 
QCD in presence
of a quark ($\mu_q$) or an isospin~\footnote{
Notice that, in order to permit a direct comparison with $\mu_q$,
we define $\mu_{\rm iso} = (\mu_u - \mu_d)/2$, where $\mu_u$ and 
$\mu_d$ are the $u$ and $d$ quark chemical potentials, i.e. a factor
2 lower than usual. 
}
 ($\mu_{\rm iso}$) chemical potential, whose 
partition function, in the standard staggered discretization 
for fermion fields, reads 
\beq
Z_{q/{\rm iso}}(T,\mu) 
\equiv \int \mathcal{D}U e^{-S_{G}} 
(\det M [\mu])^{1\over 4} 
(\det M [\pm \mu])^{1\over 4} \:,
\label{partfun1}
\eeq
where the plus/minus sign refers to the quark/isospin chemical
potential case, {$S_G$ is the lattice gauge action} and $M$ is the fermion 
matrix in the standard staggered formulation:
\begin{eqnarray}
M_{i,j} &=& a m
\delta_{i,j} + {1 \over 2} 
\sum_{\nu=1}^{3}\eta_{i,\nu}\left(U_{i,\nu}\delta_{i,j-\hat\nu}-
U^{\dag}_{i-\hat\nu,\nu}\delta_{i,j+\hat\nu}\right) \nonumber \\
&+& \frac{1}{2} \, \eta_{i,4}
\left(e^{ a \mu}\ U_{i,4}\delta_{i,j-\hat4}-
e^{- a \mu}\ U^{\dag}_{i-\hat4,4}\delta_{i,j+\hat4}\right) \, .
\label{fmatrix}
\end{eqnarray}
Here $i$ and $j$ refer to lattice sites, $\hat\nu$ is a unit vector on
the lattice, $\eta_{i,\nu}$ are the staggered phases, 
$a$ is the lattice spacing and
$m$ is the bare quark mass. 
We shall consider a bare quark mass $am = 0.05$, corresponding to 
a pion mass $m_\pi \sim 400$ MeV.

The partition function in Eq.~(\ref{partfun1}) is expressed as a functional
integral with a positive measure, hence suitable for Monte Carlo evaluation,
when $\mu$ is purely imaginary, but also when $\mu$ is real for
$Z_{\rm iso}$ alone. Our plan is to perform an extensive investigation
about the location and the nature of the deconfinement transition
for all cases in which Monte Carlo simulations are available.
The specific purposes that we have in mind are the following:\\

1)
verify the reliability of analytic continuation of the 
critical line, $T_c(\mu)$, from imaginary to 
real $\mu$
in the case of a finite $\mu_{\rm iso}$, where
simulations are available both for imaginary and real $\mu_{\rm iso}$.
Apply analytic continuation to the case of a finite $\mu_q$,
also on the basis
of what learned in the case of a finite $\mu_{\rm iso}$;\\

2)
make a careful comparison between the two theories at finite
 $\mu_q$ or $\mu_{\rm iso}$, quantifying systematic
differences for quantities like 
the curvature of the pseudocritical line at zero chemical potential;\\

3)
determine how the nature of the transition changes as a function
of the chemical potentials. In particular, in the case of a finite $\mu_q$,
no critical point is expected on the imaginary side 
since the adopted quark mass, $am = 0.05$, is 
slightly above the lower tricritical quark mass determined
for $n_f = 2$ in Ref.~\cite{rwep2}, hence the endpoint of the RW
line is second order. The situation can be different,
hence potentially more interesting,
in the case of a finite $\mu_{\rm iso}$, where the available
range of imaginary values is larger.
\\

Most numerical simulations have been performed on a $16^3 \times 4$ lattice;
different spatial sizes have been taken into account to investigate
the critical behavior in a few specific cases. The paper is organized as 
follows. In Section~\ref{sectionII} we discuss our determination of the 
critical line and its analytic continuation to real chemical
potentials. Section~\ref{curvatures} is dedicated to a systematic
comparison between the curvatures of the critical lines 
at finite $\mu_q$ and $\mu_{\rm iso}$ respectively.
In Section~\ref{order} we discuss how the order of the transition changes 
as a function of the chemical potential. 
In Section~\ref{summary} we summarize our results and draw
our conclusions. A partial account of our findings has been reported
in Ref.~\cite{proc_lat11}.

\section{Analytic continuation of the pseudocritical line}
\label{sectionII}

Before presenting results for $n_f = 2$ QCD, it 
is worth making a short summary of our previous findings.
\\

1) In SU(2) with $n_f=8$ staggered fermions and finite quark density
it was found that the analytic continuation of physical observables is 
improved if ratios of polynomials (or Pad\'e 
approximants~\cite{Lombardo:2005ks}) are used as interpolating functions~\cite{cea}.
\\

2) In SU(2) with $n_f=8$ staggered fermions and finite quark 
density~\cite{cea1,cea2} and in SU(3) with $n_f=8$ staggered fermions and 
finite isospin density~\cite{cea2} it was found that the nonlinear terms in 
the dependence of the pseudocritical coupling $\beta_c$ on $\mu^2$ in general 
cannot be neglected and that the extrapolation to real $\mu$
may be wrong otherwise.
Moreover, the coefficients of a Taylor expansion in $\mu^2$ 
of $\beta_c(\mu^2)$ were found to be all negative, 
implying subtle cancellations 
of nonlinear terms at 
imaginary $\mu$ in the first RW sector,
hence a practical difficulty in the detection of such terms 
from simulations at $\mu^2 < 0$ only.
It was realized
that, in general, a 3-parameter fit (e.g. an even polynomial of order 
$\mu^6$, with the coefficient of the $\mu^2$ term possibly constrained by 
a fit restricted to smaller $\mu^2\leq 0$ values) provided a very good 
description of the pseudocritical line in all explored cases.
\\

3) In SU(3) with $n_f=4$ staggered fermions and finite quark 
density~\cite{cea3} deviations in the pseudocritical line from the 
linear behavior in $\mu^2$ for larger absolute values of $\mu^2$ were clearly
seen. However, it turned out that several kinds of functions were able to 
interpolate them, leading to extrapolations to real $\mu$ which start 
disagreeing from each other for $\mu/T\gtrsim 0.6$. In this case, 
contrary to the studies mentioned above, direct simulations at real chemical 
potentials are not available:
one is not able to decide which extrapolation is the right one
and the disagreement is in fact a measure of the systematic ambiguity
related to analytic continuation.
\\

In the present study we approach the case of SU(3) with $n_f = 2$ and a 
standard staggered fermion discretization. In principle one  
expects that issues related to analytic continuation of the 
critical line may depend on the number of flavors, 
since the coefficients of the Taylor expansion in $\mu^2$ themselves
have such dependence: for instance it is known that 
the curvature of 
the pseudocritical line at $\mu = 0$ is smaller for smaller $n_f$, hence 
the sensitivity to nonlinear terms in $\mu^2$ could 
be enhanced. 

Moreover we shall take into account also a finite $\mu_{\rm iso}$,
considering both the imaginary
and the real potential case. That will give us the opportunity 
to directly check the validity of analytic continuation and to have
a test-ground available for the  
different extrapolations in a theory which is free of the sign problem
and is as close as possible to the one explored at finite $\mu_q$.
In view of that we shall discuss
results at finite $\mu_{\rm iso}$ at first.

The range of imaginary chemical potentials which are useful
to analytic continuation is limited by the 
periodicity in Im$(\mu)/T$, which is $2 \pi /N_c$ for $\mu_q$
and $2 \pi$ for $\mu_{\rm iso}$,
(see for instance Ref.~\cite{2im} for a discussion on this point)
and by the presence of unphysical phase transitions in the high-$T$ region.
In the explored $N_c = 3$ case, 
numerical simulations
will be limited, in the finite quark chemical potential case, 
to  Im$(\mu_q)/T \leq \pi/3$,
where the first RW transition line is met at which, in the high-$T$ region, 
the Polyakov loop switches from one $Z_3$ sector to the other. 
At finite isospin chemical potential instead we 
limit simulations to Im$(\mu_{\rm iso})/T \lesssim \pi/2$, 
where at high-$T$ a RW like transition is met at which $G$-parity is 
spontaneously broken and the Polyakov loop develops an imaginary 
part~\cite{cea2}.

We have adopted a Rational Hybrid Monte Carlo (RHMC) algorithm,
 properly modified for the inclusion
of quark/isospin chemical potential: modifications with respect
to $\mu = 0$ are trivial apart 
from the case of a real isospin chemical potential, where
the usual
even-odd factorization trick does not work 
and an additional square root of the determinant
is needed.
Typical statistics have been around 10k trajectories of 1 Molecular Dynamics
unit for each run, growing up to 100k trajectories for 4-5 $\beta$ values
around the pseudocritical point, for each $\mu^2$, 
in order to correctly sample
the critical behavior at the transition. 

The pseudocritical $\beta(\mu^2)$ 
has been determined as the value for which the susceptibility 
of the (real part of the) Polyakov loop exhibits a peak. 
To precisely localize the peak, a Lorentzian interpolation has been used.
We have verified that the determinations are consistent if the susceptibility 
of a different observable, such as the quark condensate or 
the quark number (isospin charge) is used.
In subsections~\ref{results_isospin} and \ref{results_quark} $\mu$ 
will stand respectively for $\mu_{\rm iso}$ and $\mu_q$.

\subsection{Nonzero isospin chemical potential}
\label{results_isospin}

\begin{figure}[htbp]
\includegraphics*[height=0.295\textheight,width=0.95\columnwidth]
{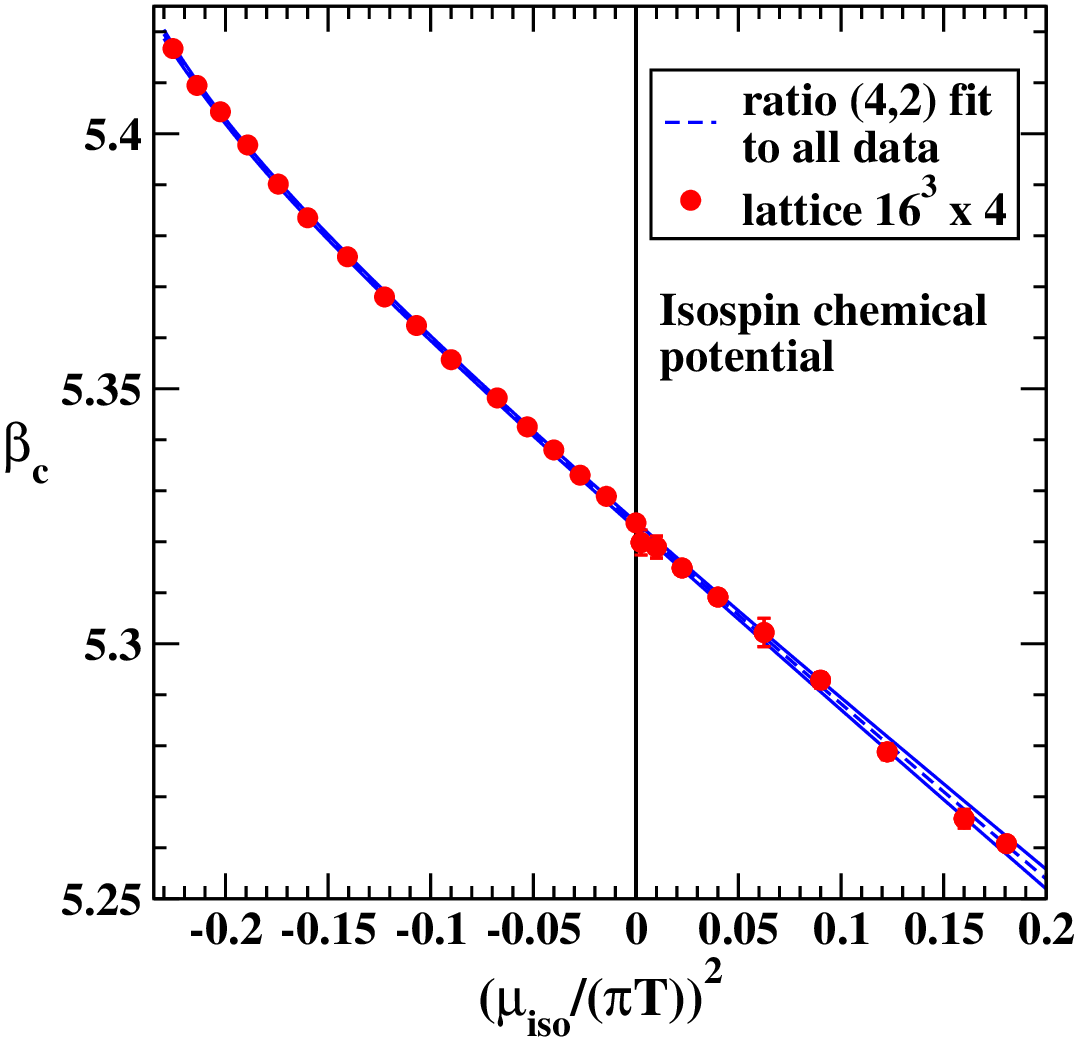}
\caption{Pseudocritical couplings obtained in finite isospin SU(3) with 
$n_f=2$ on a 16$^3\times 4$ lattice with $am$=0.05, both at $\mu^2\leq 0$
(diamonds) and $\mu^2> 0$ (circles). The dashed line represents a global
fit to all data with the ratio of a 4th- to 2nd-order polynomial; the solid
lines around the fitting curve delimit the 95\% { confidence level} (CL) 
band.}
\label{su3_isospin_beta_crit}
\end{figure}

In Table~\ref{su3_isospin_data} and in Fig.~\ref{su3_isospin_beta_crit}
we present our determinations of the pseudocritical couplings, both for 
negative and positive $\mu^2$. As a preliminary step, we have tried if an 
analytic function of $\mu^2$ exists, able to reproduce all the available data, 
both at negative {\em and} positive $\mu^2$. It turned out that no even 
polynomials in $\mu^2$ up to the 4th order can do the job and that the first 
successful global fit is achieved with a ratio of a 4th to 2nd order 
polynomial (see Fig.~\ref{su3_isospin_beta_crit} and 
Table~\ref{table_fits_isospin} for the fit parameters
and their uncertainties). 
Also the fit with a 6th-order polynomial and the  ``physical'' fit defined 
below give a reasonable global fit, with $\chi^2$/d.o.f. $\lesssim 2$.

However, it is interesting to notice that a simple linear function in $\mu^2$
fits well if one includes all data but those with $\mu^2 < -0.375^2$ (see
2nd row of Table~\ref{table_fits_isospin}). That means that, contrary to
what we observed in our previous studies, in this case non-linear corrections
are more important for imaginary values of $\mu$ than for real ones, where 
instead, in the range explored in the present study and within errors, they 
are negligible. 

\begin{table}[htbp]
\setlength{\tabcolsep}{0.5pc}
\centering
\caption[]{Summary of the values of $\beta_c(\mu^2)$ for finite isospin
SU(3) with $n_f=2$ on the 16$^3\times 4$ lattice with fermionic mass 
$am$=0.05.}
\begin{tabular}{dd}
\hline
\hline
\multicolumn{1}{c}{\hspace{0.70cm}$\mu/(\pi T)$} &
\multicolumn{1}{c}{\hspace{1cm}$\beta_c$} \\
\hline
0.475i  & 5.41670(31) \\
0.4625i & 5.40948(40) \\
0.450i  & 5.40429(51) \\
0.435i  & 5.39780(59) \\
0.4175i & 5.39012(49) \\
0.400i  & 5.38353(44) \\
0.375i  & 5.37588(61) \\
0.350i  & 5.36799(62) \\
0.327i  & 5.36239(64) \\
0.300i  & 5.35570(50) \\
0.260i  & 5.34820(47) \\
0.230i  & 5.3425(10)  \\
0.200i  & 5.33800(52) \\
0.165i  & 5.33304(85) \\
0.120i  & 5.3289(12)  \\
0.      & 5.32371(86) \\
0.050   & 5.3199(24)  \\
0.100   & 5.3189(22)  \\
0.150   & 5.31486(82) \\
0.200   & 5.3091(13)  \\
0.250   & 5.3022(28)  \\
0.300   & 5.2928(15)  \\
0.350   & 5.2788(15)  \\
0.400   & 5.2657(18)  \\
0.425   & 5.26079(94) \\
\hline
\hline
\end{tabular}
\label{su3_isospin_data}
\end{table}

We have tried several kind of interpolations of the pseudocritical couplings 
at $\mu^2 \leq 0$. 
At first, we have considered interpolations with polynomials up to
order $\mu^{6}$ (see Table~\ref{table_fits_isospin}, 3rd to 6th rows, for a 
summary of the resulting fit parameters). 
We can see that data at $\mu^2 \leq 0$ are 
precise enough to be sensitive to terms beyond the order $\mu^2$; indeed,
a good $\chi^2$/d.o.f. is not achieved before including terms up to the 
order $\mu^6$.

\begin{table*}[htbp]
\setlength{\tabcolsep}{-0.35pc}
\centering
\caption[]{Parameters of the fits to the pseudocritical couplings 
in finite isospin SU(3) with $n_f=2$ on a 16$^3\times 4$ lattice with 
fermionic mass $am$=0.05, according to the fit function
$\beta_c (\mu^2) = (a_0 + a_1 (\mu/(\pi T))^2 + a_2 (\mu/(\pi T))^4
+ a_3 (\mu/(\pi T))^6)/(1 + a_4 (\mu/(\pi T))^2)$. 
Blank columns stand for terms not included in the fit. The asterisk
denotes a constrained parameter. Fits are performed in the interval
$[\mu/(\pi T))_{\rm min}^2,\mu/(\pi T))_{\rm max}^2]$; the last two
columns give the value of $(\mu/(\pi T))_{\rm min,max}^2$.}
\begin{tabular}{dddddccc}
\hline
\hline
 \multicolumn{1}{c}{\hspace{1cm}$a_0$}
&\multicolumn{1}{c}{\hspace{1cm}$a_1$}
&\multicolumn{1}{c}{\hspace{1cm}$a_2$}
&\multicolumn{1}{c}{\hspace{1cm}$a_3$} 
&\multicolumn{1}{c}{\hspace{1cm}$a_4$} 
&\ \ \ $\chi^2$/d.o.f. \ \ \
&\ \ \ $(\mu/(\pi T))_{\rm min}^2$ \ \ \
&\ \ \ $(\mu/(\pi T))_{\rm max}^2$ \ \ \ \\
\hline
5.32326(62) & 16.755(10)   & -1.072(26) && 3.2143(19) & 0.60  & $-0.475^2$ & 
$0.425^2$ \\
5.32385(54) & -0.3597(60)  &            &            && 0.96  & $-0.375^2$ & 
$0.425^2$ \\[5pt]
5.31940(76) & -0.4192(47)  &            &            && 18.3 & $-0.475^2$ & 0\\
5.3232(11)  & -0.368(12)   &            &            && 0.59 & $-0.375^2$ & 0\\
5.3255(14)  & -0.286(25)   & 0.511(94)  &            && 1.85 & $-0.475^2$ & 0\\
5.3235(21)  & -0.374(68)   & -0.36(63)  & -2.4(1.7)  && 0.43 & $-0.475^2$ & 0\\
5.3232^*    & -0.368^*     & -0.253(91) & -2.01(44)  && 0.62 & $-0.475^2$ & 0\\
5.32403(94) & 14.602(14)   & -0.844(44) && 2.8066(25) & 0.49 & $-0.475^2$ & 0\\
\hline
\hline
\end{tabular}
\label{table_fits_isospin}
\end{table*}

As in Ref.~\cite{cea2}, we performed a ``constrained'' fit: 
first,
the largest interval $[(\mu/(\pi T))^2_{\rm min},0]$ was identified where 
data could be interpolated by a first order polynomial in $(\mu/(\pi T))^2$, 
with a $\chi^2$/d.o.f $\sim 1$; it turned out that $(\mu/(\pi T))^2_{\rm min}=
-0.375^2$. 
Then, all 
available data were fitted by a 6th-order polynomial, with the constant term 
and the quadratic 
coefficient fixed at $5.3232$ and $-0.368$, respectively
(see Table~\ref{table_fits_isospin}, 7th row). 

\begin{figure*}[tb]
\includegraphics*[height=0.295\textheight,width=0.95\columnwidth]
{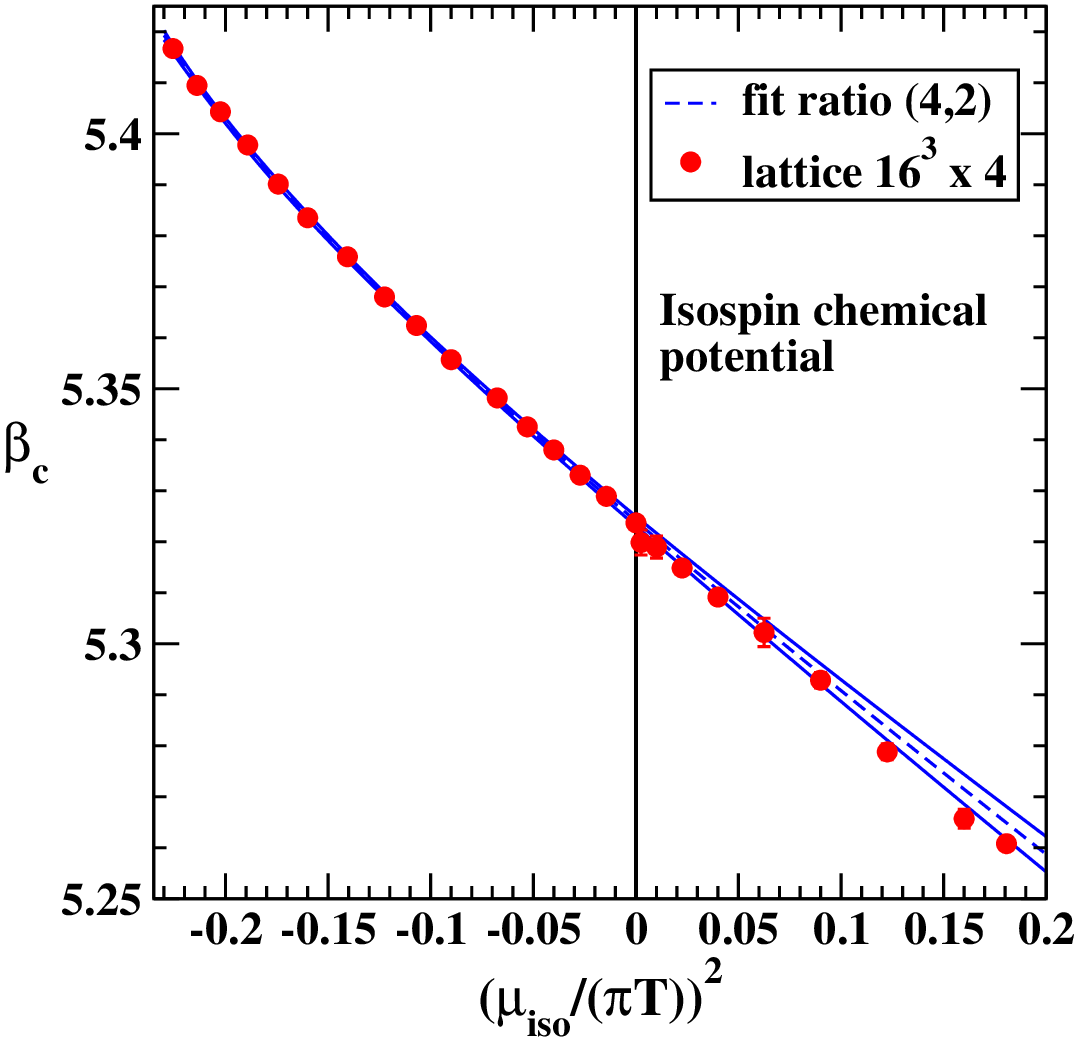}
\hspace{0.3cm}
\includegraphics*[height=0.295\textheight,width=0.95\columnwidth]
{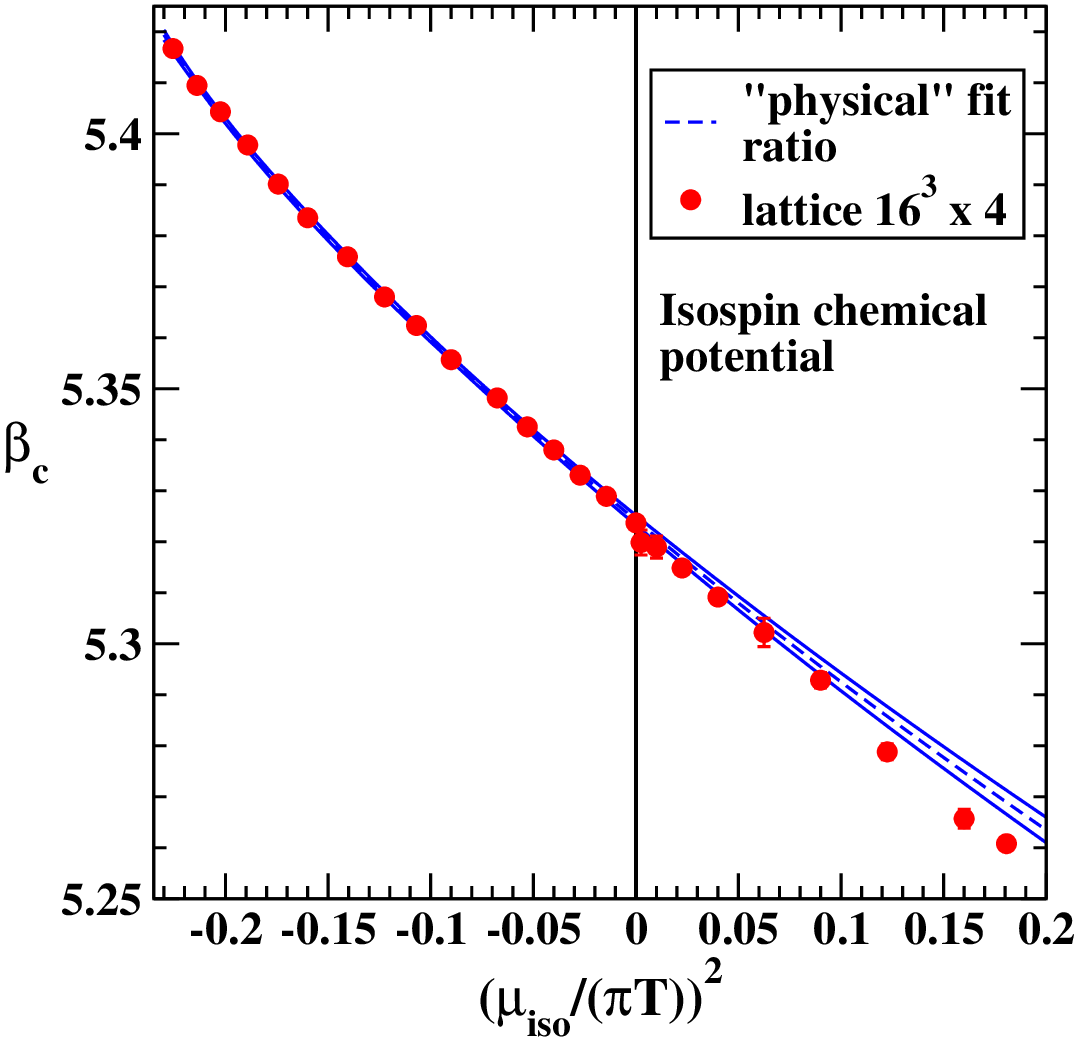}
\caption{Fits to the pseudocritical couplings in finite isospin SU(3) with 
$n_f=2$ on a 16$^3\times 4$ lattice with fermionic mass $am$=0.05:
ratio of a 4th- to 2nd-order polynomial (left) and ``physical'' fit according 
to the function~(\ref{phys_fit}) (right).} 
\label{fig_phys_fit_isospin}
\end{figure*}

Then, we have considered interpolations with ratios of polynomials of
order up to $(\mu/(\pi T))^4$. The interpolation with the least 
number of parameters for which we got a good fit is 
the ratio of a 4th- to 2nd-order polynomial, 
see Table~\ref{table_fits_isospin}, 8th row 
and Fig.~\ref{fig_phys_fit_isospin}(left). 

Finally, we have tried here the fit strategy first suggested in 
Ref.~\cite{cea2}, consisting in writing the interpolating function in 
{\em physical units} and to deduce from it the functional dependence of 
$\beta_c$ on $\mu^2$, after establishing a suitable correspondence between 
physical and lattice units. The natural, dimensionless variables of our 
theory are $T/T_c(0)$, where $T_c(0)$ is the pseudocritical temperature at 
zero chemical potential, and $\mu/(\pi T)$. The ratio $T/T_c(0)$ is deduced 
from the relation $T=1/(N_t a(\beta))$, where $N_t$ is the number of lattice 
sites in the temporal direction and $a(\beta)$ is the lattice spacing at a 
given $\beta$.
Strictly speaking 
the lattice spacing depends also on the bare quark mass, 
however in the following evaluation, 
which is only based on the perturbative 2-loop expression of
$a(\beta)$ for $N_c=3$ and $n_f=2$, we shall neglect such dependence.

We considered the following ``physical'' fit ratio
($x\equiv[\mu/(\pi T_c(\mu))]^2$):
\beq
\left[\frac{T_c(0)}{T_c(\mu)}\right]^2=\frac{1+A\,x +B\,x^2}{1+C\,x}\;, 
\label{phys_crit}
\eeq
leading to the following implicit relation between $\beta_c$ and $\mu^2$:
\beq
a^2(\beta_c(\mu^2))|_{\rm 2-loop} = a^2(\beta_c(0))|_{\rm 2-loop} 
\frac{1+A\,x + B\,x^2}{1+C\,x} \;, 
\label{phys_fit}
\eeq
with these resulting parameters
\begin{eqnarray}
\beta_c(0) &=& 5.32422(94) \;,\;\;\;\;\;\; A = 4.077(23) \;, \nonumber \\
B &=& 2.659(77) \;,\;\;\;\;\;\; C = 3.221(26) \;,
\label{phys_fit_param_isospin}
\end{eqnarray}
and $\chi^2$/d.o.f.=0.53.
In Fig.~\ref{fig_phys_fit_isospin}(right) we 
compare the ``physical'' fit ratio to data for $\beta_c(\mu^2)$. 

\begin{figure}[tb]
\includegraphics*[height=0.295\textheight,width=0.95\columnwidth]
{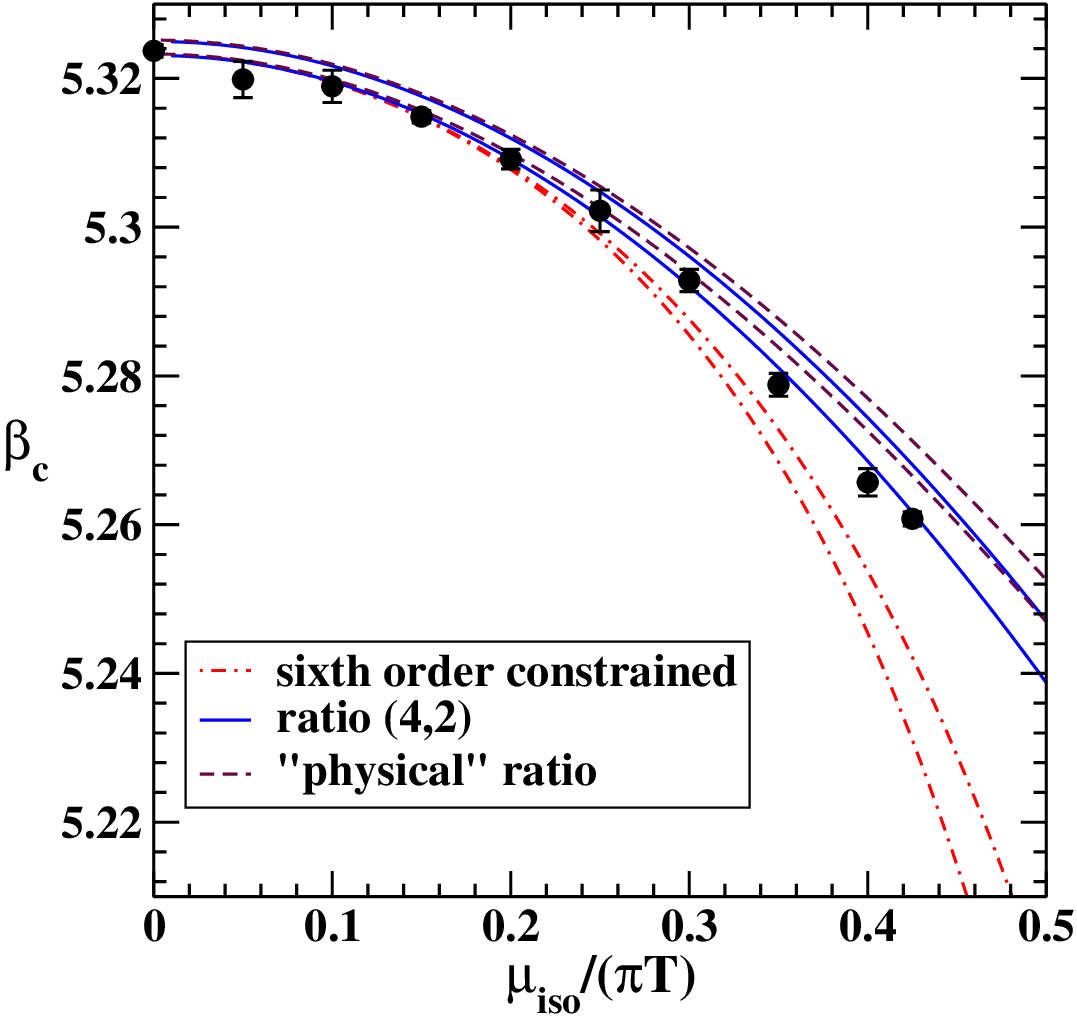}
\caption{Extrapolation to real isospin chemical potentials of the 6th-order 
constrained, ratio (4,2) of polynomials and ``physical'' ratio fits 
(only the borders of the 95\% CL band have been
reported). Data points (circles) are the results of Monte 
Carlo simulations performed directly at real isospin chemical potential.}
\label{fig_extrapolation_isospin}
\end{figure}

In Fig.~\ref{fig_extrapolation_isospin} we have plotted the 
extrapolations to the interval $0 \leq \mu/(\pi T) \leq 0.5$ of the following 
fits: 
\begin{itemize}
\item 6th-order constrained polynomial (7th row in 
Table~\ref{table_fits_isospin}); 
\item ratio (4,2) of polynomials (last row in Table~\ref{table_fits_isospin});
\item ``physical'' fit ratio, 
Eqs.~(\ref{phys_crit})-(\ref{phys_fit_param_isospin});
\end{itemize}
The three curves agree as long as $\mu/(\pi T) \lesssim 0.2$, but
then the 6th-order constrained polynomial deviates from the other two curves.
This means that different interpolations, which all reproduce
the trend of data in the fit region $-0.475^2 \leq (\mu/(\pi T))^2 \leq 0$ 
and take correctly into account the deviation from the quadratic behavior
in that region, lead to distinct extrapolations, as it occurred in $n_f=4$ 
SU(3). 
We can see that ratio of polynomials (Pad\'e approximants)
in general tend to be closer to direct determinations of the pseudocritical
couplings, which are reported in the same figure for a few values of 
$(\mu/(\pi T))^2$. 

\subsection{Nonzero quark chemical potential}
\label{results_quark}

\begin{table}[htbp]
\setlength{\tabcolsep}{0.5pc}
\centering
\caption[]{Summary of the values of $\beta_c(\mu^2)$ for finite density 
SU(3) with $n_f=2$ on the 16$^3\times 4$ lattice with fermionic mass 
$am$=0.05.}
\begin{tabular}{dd}
\hline
\hline
\multicolumn{1}{c}{\hspace{0.70cm}Im$(\mu)/(\pi T)$} &
\multicolumn{1}{c}{\hspace{1cm}$\beta_c$} \\
\hline
0.     & 5.32371(86)  \\
0.100  & 5.3277(12)   \\
0.180  & 5.33524(71)  \\
0.200  & 5.33914(83)  \\
0.245  & 5.34712(75)  \\
0.260  & 5.35000(81)  \\
0.270  & 5.35255(91)  \\
0.280  & 5.35510(59)  \\
0.290  & 5.35710(70)  \\
0.300  & 5.35970(21)  \\
0.310  & 5.36307(62)  \\
0.320  & 5.36622(37)  \\
0.327  & 5.36956(63)  \\
\multicolumn{1}{c}{\hspace{0.3cm}$1/3$} & 5.37067(75) \\
\hline
\hline
\end{tabular}
\label{su3_quark_data}
\end{table}

In Table~\ref{su3_quark_data} we summarize our determinations of the 
pseudocritical couplings.
We have tried several kinds of interpolation of the pseudocritical couplings 
at $\mu^2 \leq 0$. 
At first, we have considered interpolations with polynomials up to
order $\mu^{6}$ (see Table~\ref{table_fits_quark}, 1st to 4th row, for a 
summary of the resulting fit parameters). 
We can see that data at $\mu^2 \leq 0$ are 
precise enough to be sensitive to terms beyond the order $\mu^2$; indeed,
a good $\chi^2$/d.o.f. is not achieved before including terms up to the 
order $\mu^4$.

As in the case of isospin chemical potential, we have performed a 
``constrained'' fit. The largest interval for which a linear fit in 
$\mu^2$ works well is $[(\mu/(\pi T))^2_{\rm min},0]$ with 
$(\mu/(\pi T))^2_{\rm min}= -0.310^2$; we notice that such interval was larger 
($(\mu/(\pi T))^2_{\rm min}= -0.375^2$) in the case of an isospin chemical 
potential. Then, all available data were fitted by a 6th-order polynomial, 
with the constant term and the quadratic coefficient fixed at $5.32283$ and 
$-0.410$, respectively (see Table~\ref{table_fits_quark}, 5th row).

\begin{table*}[htbp]
\setlength{\tabcolsep}{0.4pc}
\centering
\caption[]{Parameters of the fits to the pseudocritical couplings 
in finite density SU(3) with $n_f=2$ on a 16$^3\times 4$ lattice with 
fermionic mass $am$=0.05, according to the fit function
$\beta_c (\mu^2) = (a_0 + a_1 (\mu/(\pi T))^2 + a_2 (\mu/(\pi T))^4
+ a_3 (\mu/(\pi T))^6)/(1 + a_4 (\mu/(\pi T))^2)$. 
Blank columns stand for terms not included in the fit. The asterisk
denotes a constrained parameter. Fits are performed in the interval
$(\mu/(\pi T))_{\rm min}^2,0]$; the last column gives the value of 
$(\mu/(\pi T))_{\rm min}^2$.}
\begin{tabular}{dddddcl}
\hline
\hline
\multicolumn{1}{c}{\hspace{1cm}$a_0$} & \multicolumn{1}{c}{\hspace{1cm}$a_1$} &
\multicolumn{1}{c}{\hspace{1cm}$a_2$} & \multicolumn{1}{c}{\hspace{1cm}$a_3$} &
\multicolumn{1}{c}{\hspace{1cm}$a_4$} &
$\chi^2$/d.o.f. & $(\mu/(\pi T))_{\rm min}^2$ \\
\hline
5.32189(78) & -0.4262(90) &            &            && 2.87    & $-1/3^2$   \\
5.32283(83) & -0.410(10)  &            &            && 0.63    & $-0.310^2$ \\
5.3242(13)  & -0.314(44)  & 0.92(35)   &            && 0.85    & $-1/3^2$   \\
5.3226(12)  & -0.446(86)  & -1.7(1.7)  & -14.4(9.7) && 1.41    & $-1/3^2$   \\
5.32283^*   & -0.410^*    & -0.76(13)  & -8.7(5.4)  && 0.65    & $-1/3^2$   \\
5.32394(98) & 25.736(24)  & -1.05(61)  && 4.9002(45) & 0.60    & $-1/3^2$   \\
\hline
\hline
\end{tabular}
\label{table_fits_quark}
\end{table*}

\begin{figure*}[tb]
\includegraphics*[height=0.295\textheight,width=0.95\columnwidth]
{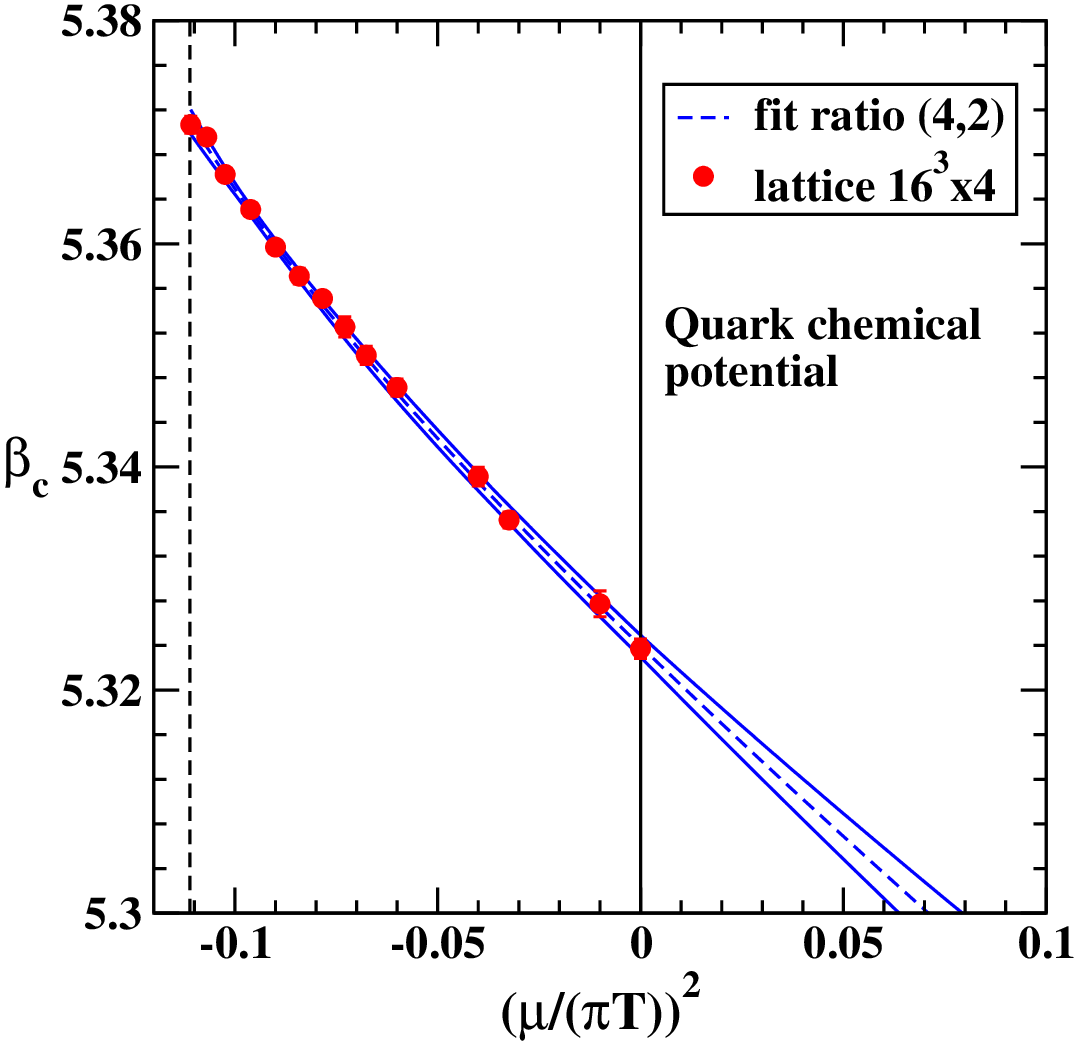}
\hspace{0.3cm}
\includegraphics*[height=0.295\textheight,width=0.95\columnwidth]
{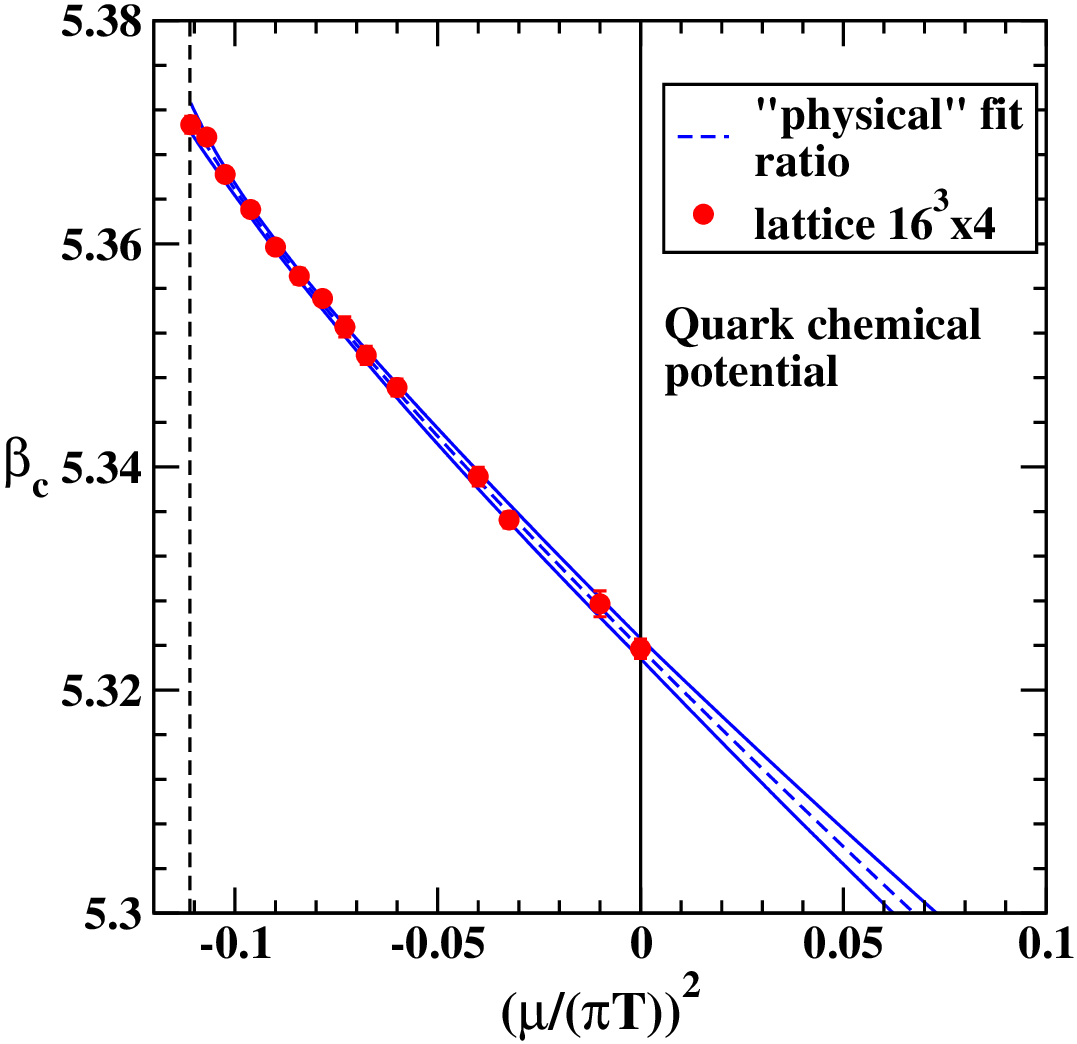}
\caption{Fits to the pseudocritical couplings in finite density SU(3) with 
$n_f=2$ on a 16$^3\times 4$ lattice with fermionic mass $am$=0.05:
ratio of a 4th- to 2nd-order polynomial (left) and ``physical'' fit according 
to the function~(\ref{phys_fit}) (right). The dashed vertical line
indicates the boundary of the first RW sector, Im$(\mu)/(\pi T)=1/3$.} 
\label{fig_phys_fit_quark}
\end{figure*}

Then, we have considered interpolations with ratios of polynomials:
the one with the 
least number of parameters for which we got a good fit
is the ratio of a 4th- to 2nd-order polynomial 
(see Table~\ref{table_fits_quark},
6th row, and Fig.~\ref{fig_phys_fit_quark}(left)).

Finally, we have also tried the ``physical'' fit, as in the 
previous subsection, obtaining the following results for the 
``physical'' fit ratio, Eq.~(\ref{phys_crit}):
\begin{eqnarray}
\beta_c(0) &=& 5.32373(90) \;,\;\;\;\;\;\; A = 8.140(32) \;, \nonumber \\
B &=& 6.59(26) \;,\;\;\;\;\;\; C = 7.201(35) \;,
\label{phys_fit_param_quark}
\end{eqnarray}
with $\chi^2$/d.o.f.=0.51, see Fig.~\ref{fig_phys_fit_quark}(right) for
a comparison of the fit to data for $\beta_c(\mu^2)$. 

Such interpolation permits, in principle, an extrapolation
to real chemical potentials down to $T = 0$; from the 
parameters given in~(\ref{phys_fit_param_quark}) 
one can get the extrapolation at $T=0$ of the pseudocritical quark chemical 
potential: $\mu_c\equiv\pi\sqrt{C/B}$ = 3.284(65) $T_c(0)$.
This result agrees within errors with the analogous one 
obtained in Ref.~\cite{NN2011} for SU(3) with $n_f=2$ Wilson fermions on a 
smaller lattice and with smaller statistics, which turned out to be 
2.73(58) $T_c(0)$.

\begin{figure}[htbp]
\includegraphics*[height=0.295\textheight,width=0.95\columnwidth]
{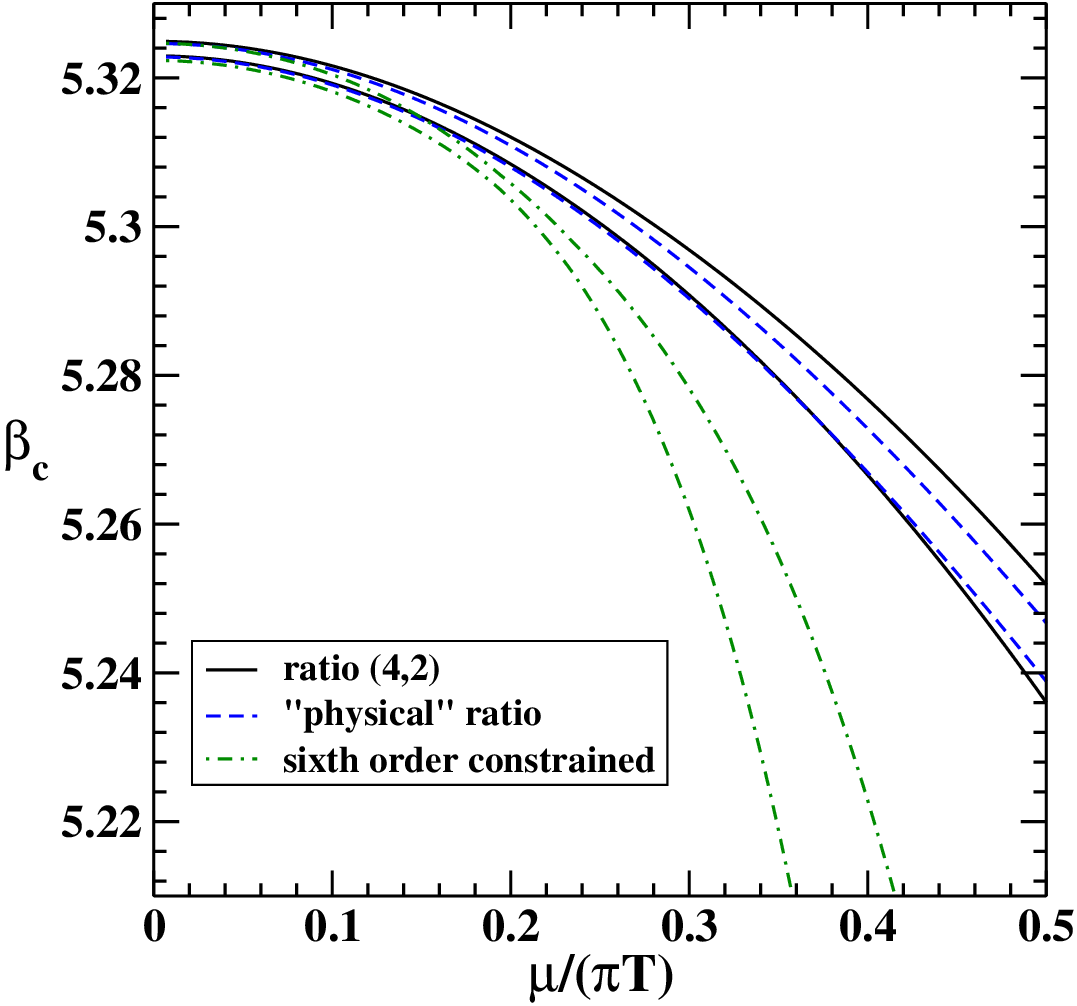}
\caption{Extrapolation to real quark chemical potentials of the 6th-order 
constrained, ratio (4,2) of polynomials and ``physical'' ratio 
fits (only the borders of the 95\% CL band have been
reported).}
\label{fig_extrapolation_quark}
\end{figure}

However, a comparison of different extrapolations shows that systematic effects
become important well before one approaches the $T = 0$ axis. 
In Fig.~\ref{fig_extrapolation_quark} we have plotted the 
extrapolations to the interval $0 \leq \mu/(\pi T) \leq 0.5$ of the following 
fits: 
\begin{itemize}
\item 6th-order constrained polynomial (5th row in 
Table~\ref{table_fits_quark}); 
\item ratio (4,2) of polynomials (last row in Table~\ref{table_fits_quark}); 
\item ``physical'' fit ratio, 
Eqs.~(\ref{phys_crit}),~(\ref{phys_fit}),~(\ref{phys_fit_param_quark}).
\end{itemize}
The three curves agree as long as $\mu/(\pi T) \lesssim 0.1$, but
then the 6th-order constrained polynomial deviates from the other two curves.
Therefore results extrapolated to larger values of $\mu$ are not reliable.
One could take the analogous results obtained at finite isospin chemical
potential as a guiding reference, concluding that Pad\'e like fits
are to be preferred; however one cannot exclude that
such argument may be wrong because of possible
systematic differences between QCD at finite quark and isospin chemical
potentials.

\section{Comparison of the curvatures of the critical lines}
\label{curvatures}

In the present Section we will focus on 
the curvature of the pseudocritical line at $\mu = 0$,  
which is the 
quantity with the least ambiguity related to the procedure 
of analytic continuation and for which a clear agreement among 
the determinations obtained by various different methods has been
shown in previous literature~\cite{krato,fodor}.
Our purpose is to 
determine how it changes when switching
from a theory at finite quark chemical potential to a theory at finite
isospin chemical potential.  

In particular we want to determine the dependence of 
the critical temperature $T_c(\mu_q,\mu_{\rm iso})$ at the quadratic
order in $\mu_q$ and $\mu_{\rm iso}$, which is determined by the two curvatures
alone. Indeed one can show that, for two degenerate flavors, the 
theory must be even under reflection of $\mu_q$ and $\mu_{\rm iso}$ separately
(see Ref.~\cite{2im} for a discussion on this point), so that the mixed
$\mu_q \mu_{\rm iso}$ term is absent and
\begin{equation}
T_c(\mu_q,\mu_{\rm iso}) = T_c(0) + A_q \mu_q^2 + A_{\rm iso} \mu_{\rm iso}^2 
+ {\cal O}(\mu_{q/{\rm iso}}^4,\mu_q^2 \mu_{\rm iso}^2) \;.
\label{curvature1}
\end{equation}

In order to determine the two curvatures and compare them in a consistent way, 
we have performed a common fit to all the pseudocritical couplings at 
imaginary potentials reported in Tables~\ref{su3_isospin_data} 
and~\ref{su3_quark_data} with the following function
\begin{eqnarray}
\beta_c(\mu_q,\mu_{\rm iso}) = \beta_c(0) 
+ a_q \left(\frac{\mu_q}{\pi T}\right)^2 
+ a_{\rm iso} \left(\frac{\mu_{\rm iso}}{\pi T}\right)^2 
\label{curvature2}
\end{eqnarray}
including as many data points, both at imaginary and real (when available) chemical potentials, 
as compatible with a reasonable value of 
$\chi^2$/d.o.f.
As a matter of fact the ranges of included chemical potentials
coincide with those for which a linear fit works well separately for
$\mu_q^2$ or $\mu_{\rm iso}^2$ (see the second row of Tables~\ref{table_fits_isospin}
and \ref{table_fits_quark}), the only difference in this case being that 
$\beta_c(0)$ is taken as a common parameter. The results of the fit are
\begin{eqnarray}
a_q &=& -0.3997(87) \;, \;\;\; a_{\rm iso} = -0.3606(67) \nonumber \\
\beta_c(0) &=& 5.32370(57) \;, \;\;\; \chi^2/{\rm d.o.f.} = 0.93\;.
\label{commonfit}
\end{eqnarray}
We notice that $a_q$ and $a_{\rm iso}$
are not compatible within errors and deviate from each other by about 
4$\sigma$. 

A convenient way to report the two curvatures is in term of dimensionless 
quantities, as follows:
\begin{equation}
\frac{T_c(\mu_q,\mu_{\rm iso})}{T_c(0)} = 1 
+ R_q \left(\frac{\mu_q}{\pi T}\right)^2 
+ R_{\rm iso} \left(\frac{\mu_{\rm iso}}{\pi T}\right)^2 \, .
\label{curvature3}
\end{equation}
The parameters $R_q$ and $R_{\rm iso}$ can be obtained respectively from 
$a_q$ and $a_{\rm iso}$, in particular one has
\begin{eqnarray}
R_{q/{\rm iso}} &=& 
\left. - \frac{1}{a} \frac{\partial\ a}{\partial \beta} \right|_{\beta_c(0)} 
a_{q/{\rm iso}} \nonumber \\ &=& 
\sqrt{\frac{N_c}{2 \beta_c(0)^3}} 
\frac{1}{\beta_L(\beta_c(0),m_q)} a_{q/{\rm iso}}\;,
\end{eqnarray}
where $a$ is the lattice spacing and $\beta_L = a (\partial g_0/\partial a)$ 
is the lattice beta-function. Making use of the perturbative two-loop 
expression for $\beta_L$, we get 
\begin{equation}
R_q = -0.515(11) \;,\;\;\; R_{\rm iso} = -0.465(9) \;.
\end{equation}

It is interesting to compare our results with those of previous studies.
In Ref.~\cite{kogut_curvature} the same discretization
and bare quark mass have been adopted for QCD at real isospin chemical
potential; their result, when reported in the same units
as ours, is $R_{\rm iso} = 0.426(19)$: the marginal discrepancy can
be explained in terms of either the inexact R-algorithm or the smaller
spatial volume used 
in Ref.~\cite{kogut_curvature}.
$R_q = -0.500(34)$ has been obtained in Ref.~\cite{muim} for the 
same theory with a smaller fermion mass, $am=0.025$: this is compatible with 
our result, showing that $R_q$ has mild dependence on the quark mass.
In Ref.~\cite{NN2011} a value $R_q = -0.38(12)$ has been reported making
use of $n_f = 2$ Wilson fermions: the agreement, even if within quite large errors,
is encouraging if we consider the completely different fermion discretization.
Instead, as it is well known, 
the curvature changes significantly if we change the number of 
flavors; for instance for $n_f=4$ QCD one obtains 
$R_q = -0.792(10)$~\cite{immu_dl,cea3}.

\begin{figure}[htb]
\includegraphics*[height=0.295\textheight,width=0.95\columnwidth]
{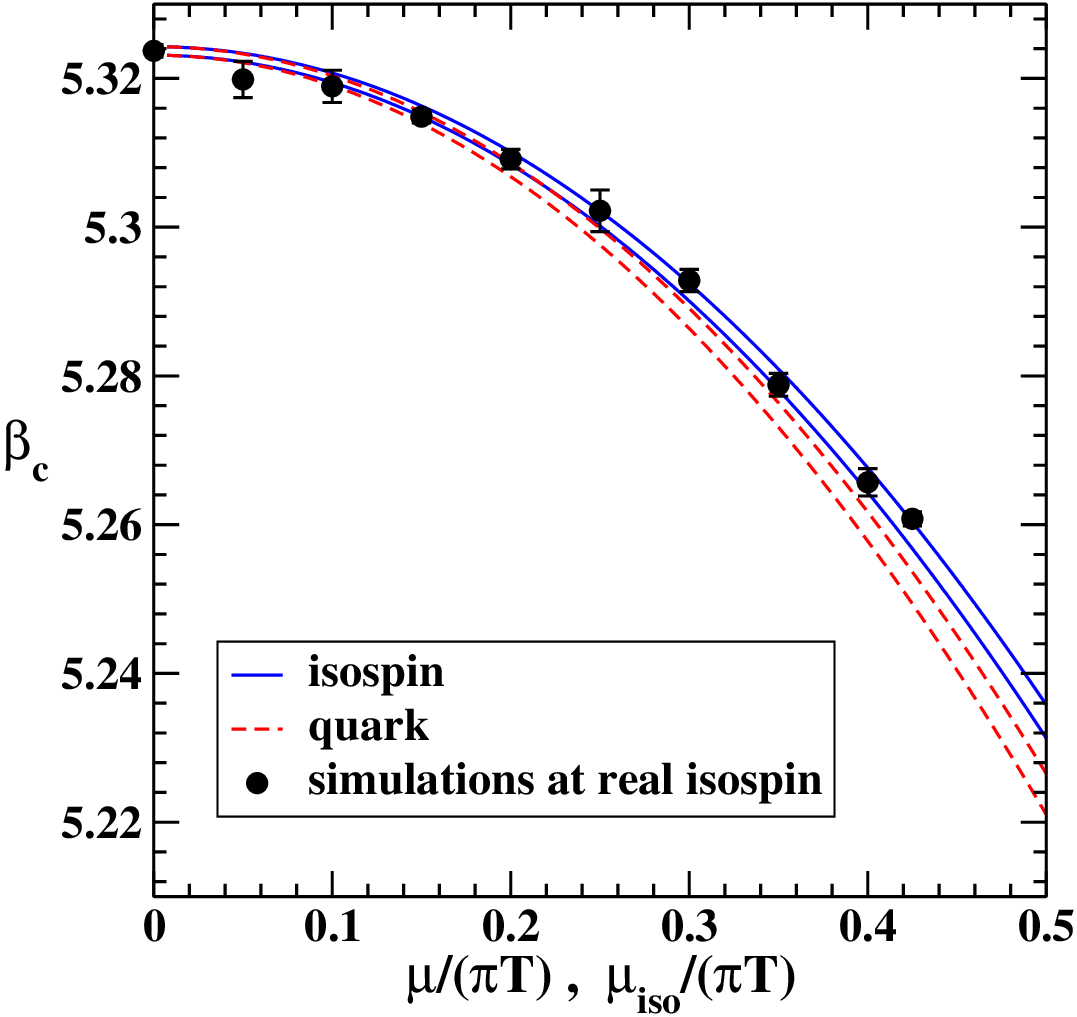}
\caption{Comparison between the extrapolations to real quark and isospin 
chemical potential of the fits linear in $\mu/(\pi T)^2$. Data points
(circles) are the results of Monte Carlo simulations performed directly 
at real isospin chemical potential.} 
\label{fig_curvature}
\end{figure}

Our determinations of $R_q$ and $R_{\rm iso}$ are clearly affected
by the systematic error related to the choice of the two-loop
expression for $\beta_L$, anyway such 
error disappears if we consider the ratio
\begin{equation}
R_{q - {\rm iso}} = \frac{R_q - R_{\rm iso}}{R_q} = \frac{a_q - a_{\rm iso}}
{a_q} = 0.098(26) \;,
\end{equation} 
which we consider as our final estimate for the difference in the curvature of the 
critical line between the theory at finite baryon density and the theory at finite isospin density.
In order to appreciate the difference, in Fig.~\ref{fig_curvature} 
we report the corresponding linear extrapolations to real chemical potentials.

In previous studies the two curvatures revealed to be 
compatible within errors~\cite{taylor1,kogut_curvature}. This is also
the expectation in the limit of a large number of colors 
$N_c$~\cite{toublan,largen1,largen2,largen3}: indeed the two curvatures are 
expected to be the same at the leading order $1/N_c$~\cite{toublan} 
(the curvature itself is expected to vanish as $N_c \to \infty$).
Therefore, we can consider the deviation that we find as the first
evidence for an $O(1/N_c^2)$ difference between the two theories 
at small chemical potentials. 
$R_{q - {\rm iso}}$, being the ratio of an  $O(1/N_c^2)$ to
an  $O(1/N_c)$ quantity, is expected to be $O(1/N_c)$: this is compatible
with the fact that it turns out to be 
of the order of 10\%. It would be interesting to 
explore how results change for different values of $N_c$.

\section{Order of the phase transition}
\label{order}

As already stressed in the Introduction, 
the nature of the pseudocritical line at imaginary
$\mu_q$, may be strongly influenced by the 
order of the RW endpoint~\cite{rwep,rwep_nf3,rwep2,rwep3}, {\it i.e.} the 
point at which the RW line taking place in the high-$T$ region
for Im$(\mu_q)/T = \pi/3$ meets the
analytic 
continuation of the physical pseudocritical line. If the endpoint
is first order then it is actually a triple point and at least the 
part of the pseudocritical line which is closest to the endpoint 
is expected
to be first order. 

In the case of $n_f = 2$, with the same regularization and temporal size ($N_t = 4$)
used in the present study,
it is known that the RW endpoint is first order, in the low mass region, for 
$am < am_{t1}$ with $am_{t1} = 0.043(5)$~\cite{rwep2}. 
That means that the mass
used in the present work, $am = 0.05$, is close to the tricritical value
but slightly on the second order side, so we do not expect the analytic
continuation of the pseudocritical line to become first order as we approach
the RW endpoint. This is compatible with the fact that we have not observed
signals of metastable behavior or double peak distributions
along the line; only a strengthening of the transition
can be seen as the RW endpoint is approached, as a consequence of the 
closeness of the tricritical point.

If one conjectures that an imaginary $\mu_{\rm iso}$ may
strengthen the transition in the same way as an imaginary $\mu_q$ does,
then, since the range available for Im$(\mu_{\rm iso})$ is larger than 
that available for Im$(\mu_q)$, one may expect that a first order
transition could be manifest at some stage along the 
pseudocritical line at imaginary $\mu_{\rm iso}$.
Such conjecture is well founded, since simulations at real 
isospin chemical potential have shown that indeed the effect of small
positive values of $\mu_{\rm iso}^2$ is a weakening of the 
transition~\cite{wenger,kogut_3flv}. 

\begin{figure}[t!]
\includegraphics*[height=0.25\textheight,width=0.95\columnwidth]
{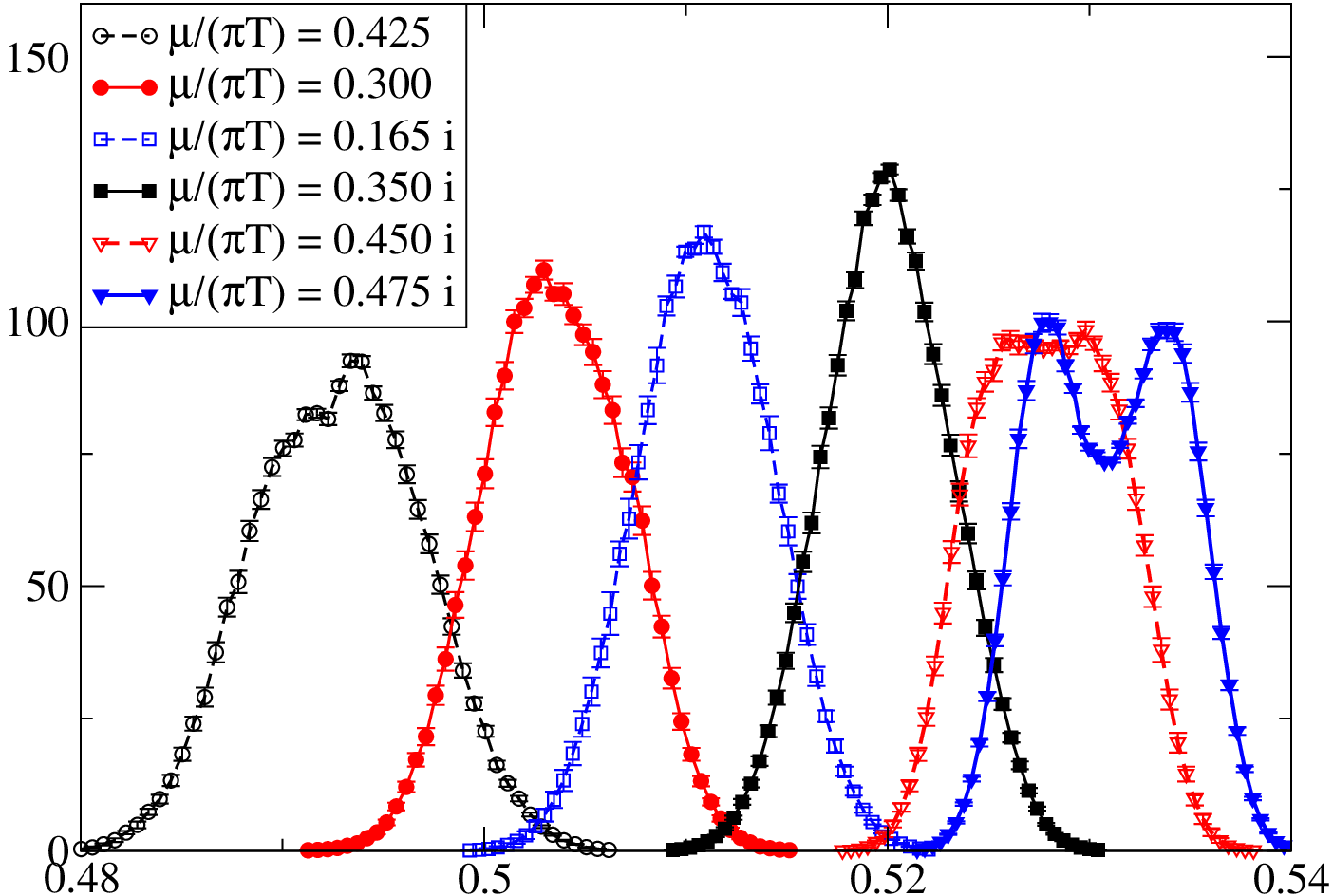}
\caption{Normalized plaquette distributions at the pseudocritical coupling for different
values of the isospin chemical potential.
}
\label{histo_L16}
\end{figure}

In order to explore this possibility, we have reported in Fig.~\ref{histo_L16}
the plaquette distributions at the pseudocritical coupling for a few different
values (both real and imaginary) of $\mu_{\rm iso}/(\pi T)$. It is evident that
for the largest values of $\mu_{\rm iso}$ a double peak structure develops, 
hinting at the presence of a first order transition. 

\begin{figure}[htb]
\includegraphics*[height=0.25\textheight,width=0.95\columnwidth]
{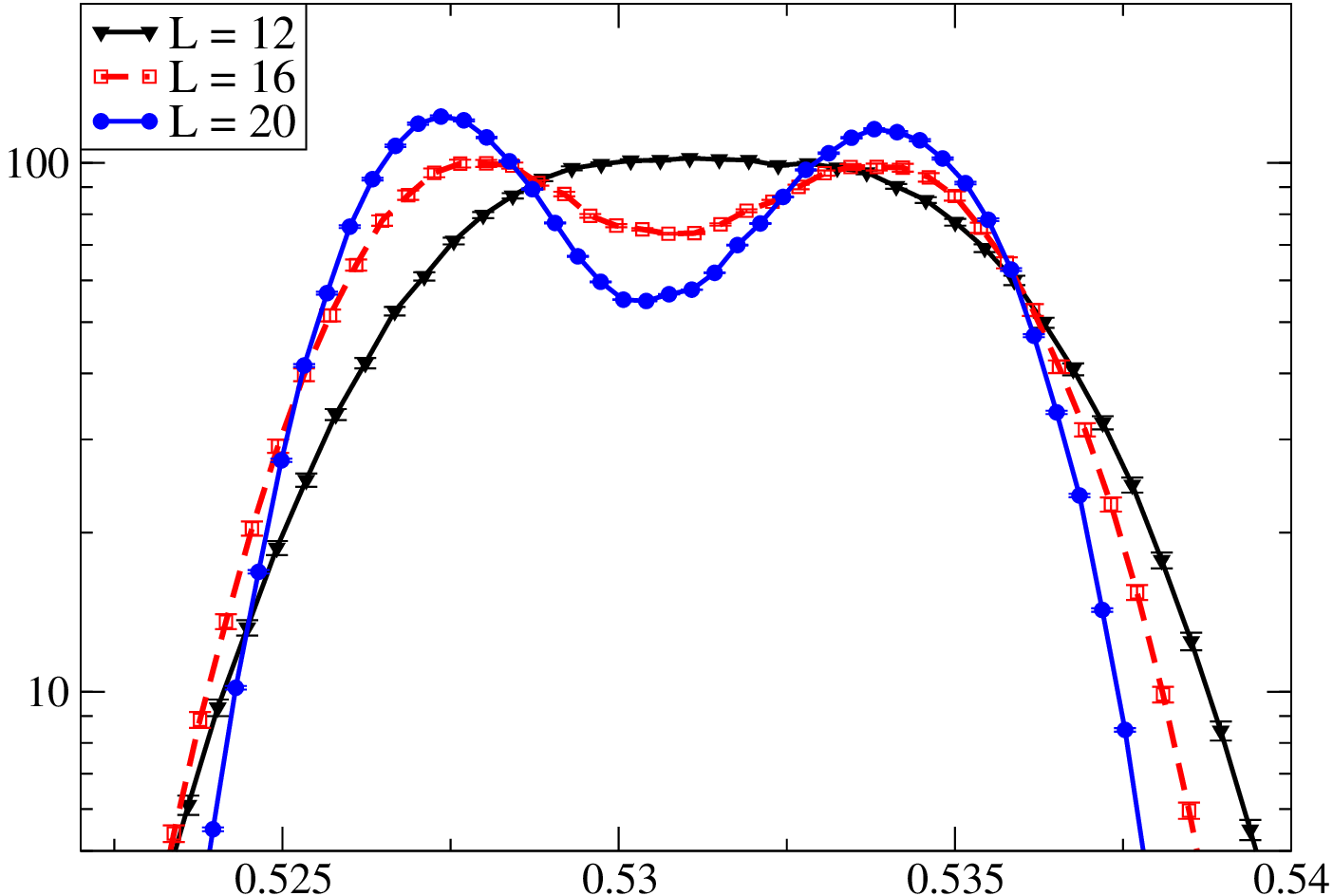}
\caption{Normalized plaquette distributions 
at the pseudocritical coupling for different
spatial lattice sizes and $\mu_{\rm iso}/(\pi T) = 0.475i$.
}
\label{histo_0.475}
\end{figure}

\begin{figure}[htb]
\includegraphics*[height=0.25\textheight,width=0.95\columnwidth]
{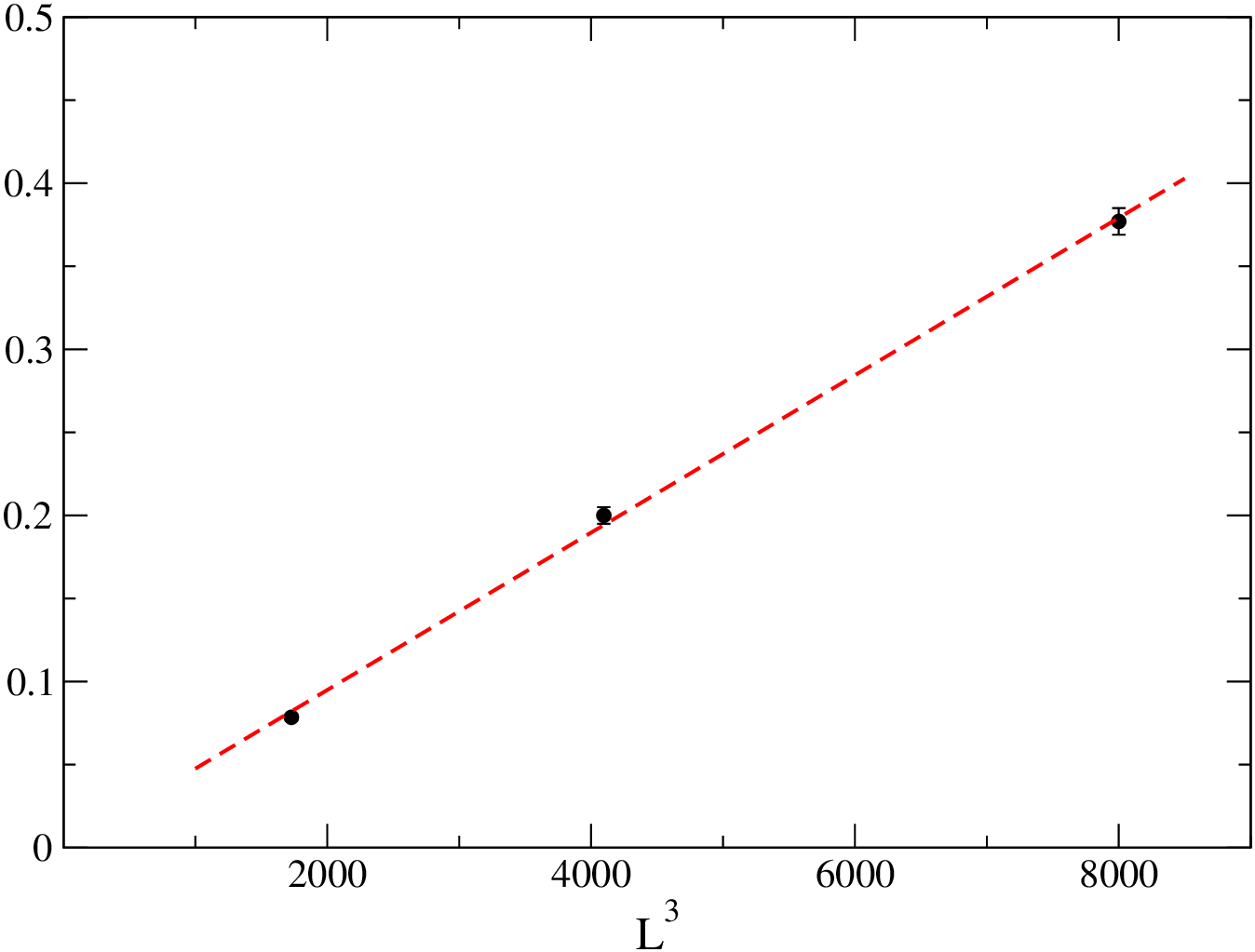}
\caption{Maxima of the plaquette susceptibility as a function of the 
spatial volume for $\mu_{\rm iso}/(\pi T) = 0.475i$.}
\label{suscmax}
\end{figure}

In order to confirm that by a finite size scaling analysis, 
we have repeated simulations for 
the largest value of Im$(\mu_{\rm iso})$,
$\mu_{\rm iso}/(\pi T) = 0.475i$, on two other lattice
sizes, $L = 12$ and $L = 20$. Both the scaling of distributions and the scaling
of susceptibilities confirm the first order nature of the transition for
this value of $\mu_{\rm iso}$: the well in the double peak distribution
of the plaquette deepens as $L$ increases as expected (see Fig.~\ref{histo_0.475})
and the maxima of the plaquette susceptibility scale linearly with the spatial
volume (see Fig.~\ref{suscmax}).

Therefore we conclude that, for the present discretization and 
value of the quark mass, the transition is surely first order 
at $\mu_{\rm iso}/(\pi T) = 0.475i$ and there is possibly 
a critical point along the line at some smaller value of 
Im$(\mu_{\rm iso})$. Such non-trivial behavior resembles what happens 
for quark chemical potentials~\cite{rwep,rwep_nf3,rwep2,rwep3} and 
may have consequences on the general 
structure of the QCD phase diagram which should be further investigated
in the future.

\section{Conclusions}
\label{summary}

In this paper we have considered QCD with two degenerate flavors of bare mass
$am=0.05$, corresponding to a pion mass $m_\pi \sim 400$ MeV, on a 
$16^3\times4$ lattice, at nonzero quark or isospin density.

Our investigation developed along three main lines:
\begin{itemize}
\item localization of the pseudocritical line in the temperature-chemical
potential plane, for the two cases of quark and isospin density;
\item comparison of the curvatures of the two critical lines at the point
of zero chemical potential;
\item study of the order of the phase transition along the two critical lines.
\end{itemize}

To study the localization of the critical line we adopted the method of 
analytic continuation. The result we got is similar to what we found in the
case of four degenerate flavors and is common to the case of quark and 
isospin density: deviations from the linear behavior in $\mu^2$
of the critical lines are clearly seen for $\mu^2<0$ and are nicely 
described by several analytic functions. However, the extrapolations to 
positive $\mu^2$ overlap, within errors, only 
as long as $\mu_{\rm iso}/(\pi T)\simeq 0.2$ and $\mu_q/(\pi T)\simeq 0.1$.
The comparison with direct numerical simulations performed at real
isospin chemical potentials leads to a preference for extrapolations
based on Pad\'e approximants; such suggestion, which is agreement
with our previous studies, could be taken as 
a guiding principle also in the case of 
nonzero quark density.

We have performed a careful determination of the curvatures of the two 
critical lines at zero chemical potential. In order to compare 
them in a consistent way, we have performed a common fit of the pseudocritical 
couplings, taking the critical $\beta$ at zero chemical potential 
as a common parameter. We have found that the curvature of the isospin critical
line is larger 
than that of the quark critical line by about 4$\sigma$, the
relative difference being about 10\%.
The outcome of previous studies~\cite{taylor1,kogut_curvature} was
in favour of a substantial agreement between the two curvatures,
as expected 
in the limit of a large number of colors 
$N_c$~\cite{toublan,largen1,largen2,largen3} and in particular 
at the leading $1/N_c$ order~\cite{toublan}. 
The deviation that we find is therefore a first
evidence for an $O(1/N_c^2)$ difference between the two theories 
at small chemical potentials. The order of magnitude of the
relative deviation
$R_{q - {\rm iso}}$, which is the ratio of an  $O(1/N_c^2)$ to
an  $O(1/N_c)$ quantity (the curvature itself)
is compatible with it being an $O(1/N_c)$ quantity.
It would be interesting to 
explore how results change for different values of $N_c$.

Finally, we have studied the order of the transitions along the critical lines.
For the case of nonzero quark density, we have found no clear 
signatures of a first 
order transition, in agreement with the expectations after recent findings in
the literature
and in consideration of the quark mass adopted in this work, which 
is larger than the tricritical mass found in Ref.~\cite{rwep2}.
The only effect we could see was a strengthening of the transition when the 
RW point is approached at imaginary quark chemical potential.

A phenomenon emerged from our investigation is instead that an 
imaginary isospin chemical potential can strengthen the transition,
similarly to what happens for a quark chemical potential and in agreement
with the fact that simulations at real isospin chemical
isospin have shown that small
positive values of $\mu_{\rm iso}^2$ weaken the 
transition~\cite{wenger,kogut_3flv}.
Moreover 
we have found clear evidence that, in this case, the
transition becomes first order for large enough imaginary chemical potentials,
but before one reaches the RW-like transition which is found,
in the high-$T$ region, for $\mu_{\rm iso}/(\pi T) \simeq 0.5$~\cite{cea2}, 
implying a 
possible second order critical point along the line. 
Such behavior is very similar to what found for a quark chemical potential,
hinting at a common underlying physical mechanism, and 
could have various non-trivial consequences, on the shape
of the critical line and on the general structure of the 
QCD phase diagram, that should be further investigated in future studies.

\section{Acknowledgments}
We are grateful to Claudio Bonati, Philippe de Forcrand and Owe Philipsen
for useful discussions. 
We acknowledge the use of the PC clusters of the INFN Bari Computer 
Center for Science and of the INFN-Genova Section.

\end{document}